\documentclass{aastex63}
\usepackage{amsmath}
\usepackage{amssymb}

\accepted{April 26, 2021}

\shorttitle{Kepler-221 Tidal Origin}

\begin{document}
\title{A Tidal Origin for a 3-body Resonance in Kepler-221}

\correspondingauthor{Max Goldberg}
\email{mg@astro.caltech.edu}

\author[0000-0003-3868-3663]{Max Goldberg}
\affiliation{Department of Astronomy, California Institute of Technology\\1200 E. California Blvd\\Pasadena, CA 91125, USA}

\author[0000-0002-7094-7908]{Konstantin Batygin}
\affiliation{Division of Geological and Planetary Sciences, California Institute of Technology\\1200 E. California Blvd\\Pasadena, CA 91125, USA}

\shortauthors{Goldberg \& Batygin}

\begin{abstract}
Over the course of the last two decades, traditional models of planet formation have been repeatedly challenged by the emerging census of extrasolar planets. Key among them is the orbital architecture problem: while standard models of orbital migration predict resonant orbits for short-period objects, most planets do not appear to lie in orbital resonances. Here we show that the four-planet system Kepler-221, not previously recognized to have active orbital resonances, has a three-body commensurability relation unique within the Kepler sample. Using a suite of numerical experiments as well as a perturbative analysis, we demonstrate that this system likely began as a resonant chain and proceeded to undergo large-scale divergence away from resonance, under the action of tidal dissipation. Our results further indicate that obliquity tides, driven by a secular spin-orbit resonance and mutual inclination, are an excellent candidate for driving this orbital divergence, and that the high tidal luminosity may also explain the anomalous size of planet b, which lies within the Fulton radius gap.
\end{abstract}


\section{Introduction}

The tally of extrasolar planets has dramatically increased over the past decade and a half. Transit and radial velocity surveys have uncovered hundreds of exoplanetary systems with multiple planets and architectures differing considerably from our Solar System. Accordingly, the census of extrasolar planets with orbital periods shorter than a ~year has come into unprecedented level of focus \citep{Fabrycky2014}. 

Among the key overall results that has stemmed from the recent flurry of planetary detections has been a characterization of ``conventional'' orbital architecture displayed by close-in extrasolar planets. At face value, the properties of a representative short-period planetary system are easily summarized: \deleted{stereo}typical planets have sizes that are a few times larger than that of the Earth, often occur in multiples, and occupy nearly planar, circular orbits that are separated by tens of mutual Hill radii \citep{Hadden2014a, Weiss2018, Millholland2017, Wang2017}. Furthermore, most multi-planet systems exhibit long-term ($\sim$Gyr) dynamical stability \citep{Tamayo2020}.

In exceptional cases, important additional insight into the orbital machinery of a given system can be gleaned from combined transit timing analysis and radial velocity measurements \citep{Petigura2018, Petigura2020}. Direct characterization aside, however, a thornier question concerns how orbital layouts of extrasolar planets arise in the first place. In this regard, a complete understanding of the physical processes that shape exoplanetary configurations remains elusive, and constitutes a topic of active research (see \cite{Raymond2020} for a review). In fact, even the epoch at which the final orbital architecture is set is a subject of debate.

Within the context of ``classical'' models of planet formation, the present-day architecture of planetary systems was assumed to be inherited from the detailed structure of the protoplanetary disk from which the planets emerged \citep{Cameron1988}. Recent theoretical progress on the origin of the exoplanetary period distribution, as well as the early evolution of the solar system itself, however, indicates that violent, post-nebular dynamical instabilities may play an important role in sculpting the terminal outcome of the planet formation process \citep{Tsiganis2005, Morbidelli2010, Izidoro2017, Esteves2020}. Unfortunately, this markedly chaotic evolutionary framework renders a deterministic model that unambiguously connects the properties of the planets’ natal disk with their present-day (observable) attributes an impossibility.

A somewhat rare exception to this rule of thumb are resonant chains---systems of planets locked into sequential mean-motion commensurabilities with one-another. Because resonant entrainment necessitates dissipative, extrinsically-driven evolution of planetary orbits, systems that exhibit resonant dynamics today are routinely interpreted as signposts of convergent orbital migration that was facilitated by their natal disks. In other words, resonant chains represent ``pristine'' orbital architectures, sculpted by the birth environment of the planets. Examples of such systems include GJ876, Kepler-80, Kepler-223, and TRAPPIST-1, and it is worth noting that dynamical origins of each of these systems have been extensively studied extensively within the literature \citep[][and references therein]{Rivera2010, MacDonald2016, Mills2016, Gillon2017, Agol2020}. \added{Precise characterization of the resonant configuration has even been used to constrain the migration rate and order of capture into resonance \citep{Delisle2017}.} 

Even more unique within the current census of exoplanets is the orbital structure of the Kepler-221 system. Unlike the vast majority of known multi-planetary configurations, Kepler-221 is not devoid of orbital commensurability. Simultaneously, however, Kepler-221 is not a standard example of a resonant chain. Rather, it exhibits a three-body commensurability far away from any discernible two-body resonances. Understanding the genesis and long-term evolution of this remarkable aggregate of planets is the primary goal of this work. In particular, here we demonstrate that the current configuration of Kepler-221 is unlikely to have formed in-situ. Instead, it can be readily understood as a resonant chain that underwent long-range divergent migration as a consequence of persistent tidal damping. Moreover, we argue that tidal dissipation in this system almost certainly stems from obliquity tides, implying that the spin axes of at least one of the Kepler-221 planets are significantly misaligned with respect to the orbital plane.

The remainder of the manuscript is organized as follows.  In Section 2, we briefly review orbital resonances.  We introduce Kepler-221 in Section 3 in the context of other Kepler multiplanet systems.  In Sections 4 and 5 we evaluate two potential mechanisms for forming the resonances in Kepler-221.  Finally, in Section 6 we summarize our results, delineate the limitations of our work and discuss the value of future constraints.

\section{Orbital Resonances} \label{sec:resonances}

Orbital mean motion resonances (MMRs) are defined by the libration (bounded oscillation) of a critical \textit{resonant} angle. For a first-order (in eccentricity) two-body resonance, the resonant angle takes the form
\begin{equation}
    \theta = p\lambda - (p+1)\lambda' + \varpi
\end{equation}
where $\lambda$ and $\lambda'$ are the mean longitudes of the inner and outer body, respectively, $\varpi$ is the longitude of pericenter for either of the bodies, and $p$ is an integer. Because apsidal precession is typically much slower than orbital motion ($|\dot{\varpi}| \ll n$), the associated period ratio is $P'/P\sim (p+1)/p$. For systems with more than two planets, critical angles can be combined to yield three-body relations. 
For example, in the Kepler-223 system mentioned above, the angles
\begin{eqnarray}
    \phi_1 = \lambda_b + 2\lambda_c - \lambda_d \\
    \phi_2 = \lambda_c - 3\lambda_d + 2\lambda_e
\end{eqnarray}
librate. Such relations are of considerable importance because transit timing variation (TTV) analyses of some resonant systems have shown that \textit{only} these angles librate \citep{Gozdziewski2016, Mills2016, MacDonald2016}. 
The formation of these systems (where three-body angles librate but two-body angles do not) has perplexed previous studies and remains an unanswered question. The aforementioned works have generally relied on the possibility that TTV analyses lack sufficiently precise eccentricity vector data to demonstrate that two-body angles are librating, and that further data would show that they in fact are librating. However, the Kepler-221 system, the subject of this study, indicates that three-body resonances can be active very far from two-body resonances. Following \cite{Gozdziewski2016}, we denote systems with librating three-body resonant angles, but no librating two-body resonant angles, \textit{pure} three-body resonances.

Kepler-221 (also known as KOI-720) is a G-type star ($T_\text{eff} = 5255$K) in the Kepler field. The Kepler data reduction pipeline identified four planets with the parameters shown in Table \ref{params}. The innermost planet, b, has a radius near the radius gap between super-Earths and sub-Neptunes, whereas the outer three planets have radii in excess of $2 R_\oplus$, placing them firmly in the sub-Neptune category \citep{Fulton2017}. The period ratios of adjacent planets, starting with c to b, are 2.035, 1.765, and 1.829. With the exception of b and c, which appear to be a few percent wide of a 2:1 resonance, none of the planet pairs lie near first- or second-order two-body commensurabilities. This masses of the planets, are, unfortunately, unconstrained.

Despite the lack of pronounced two-body resonances within the system, a three-body commensurability exists \citep{Fabrycky2014}. Specifically, the frequency
\begin{equation}
    B = 2n_b - 5n_c + 3n_e \approx -0.000727 \pm 0.000440 \text{ degrees/day}
    \label{eq:B}
\end{equation}
is exceptionally small, suggesting possible libration of the critical angle
\begin{equation}
    \phi = 2\lambda_b - 5\lambda_c + 3\lambda_e.
\end{equation}
If true, the orbital clockwork exhibited by these planets would render Kepler-221 a genuinely unusual member of the Kepler planetary census, and entail remarkable constraints on its long-term tidal evolution.

Beyond a peculiar orbital architecture, Kepler-221 exhibits markers of exceptional youth within the Kepler sample. In particular, stellar lithium abundance is an age diagnostic because Li is rapidly consumed in fusion reactions relatively early in a star's lifecycle. While determining precise age from lithium features is difficult, Kepler-221's large lithium abundance is a strong indication that it is younger than the Hyades ($\sim 650$ Myr) \citep{Berger2018}. Additionally, \replaced{its chromospheric activity index, which measures calcium emission in the chromosphere relative to the bolometric luminosity of the star, and is a common proxy for age, is $\log{R'_\text{HK}}=-4.49$, comparable with stars in the Hyades and consistent with an age of $\sim 600$ Myr \citep{Mamajek2008}}{the California-Kepler Survey \citep{Petigura2017} obtained a high-resolution optical spectrum of Kepler-221 using Keck-HIRES to measure bulk stellar properties. Based on an analysis of that spectrum using the technique of \cite{Isaacson2010}, the California Planet Search team (Andrew Howard, private communication) found a stellar activity metric of $\log R'{\mbox{\scriptsize HK}} = -4.49$, implying a rough stellar age of $\sim600$ Myr comparable to the Hyades cluster \citep{Mamajek2008}}. In contrast with this estimate, many studies of orbital evolution of Kepler planets through tidal dissipation assume Sun-like ages of 5 or 10 Gyr \citep{Silburt2015, Lee2013,Millholland2019}. As we will show below, this order-of-magnitude discrepancy in evolutionary timescale translates to strong constraints on tidal parameters and the dissipation mechanism of Kepler-221's planets.

\begin{deluxetable*}{ccccc}
\tablecaption{Observed transit parameters of the Kepler-221 system.\label{params}}
\tablewidth{0pt}
\tablehead{
\colhead{Planet} & \colhead{Period} & \colhead{Transit mid-point} & \colhead{Radius} & \colhead{Impact parameter}\\
& \colhead{(d)} & \colhead{(BJD-2454900)} & \colhead{($R_\oplus$)} &
}
\startdata
    b & $2.795906\pm 0.000004$ & $65.72929 \pm0.00084$ & $1.71\pm0.17$ & $0.61\pm0.24$ \\
    c & $5.690586\pm 0.000004$ & $107.04865\pm0.00037$ & $2.93\pm0.27$ & $0.04\pm0.16$ \\
    d & $10.041560\pm0.000011$ & $70.08456 \pm0.00060$ & $2.73\pm0.25$ & $0.36\pm0.20$ \\
    e & $18.369917\pm0.000029$ & $64.86048 \pm0.00087$ & $2.63\pm0.25$ & $0.26\pm0.22$ \\
\enddata
\end{deluxetable*}

\subsection{Resonant libration in the Kepler-221 system?}
Measuring the libration of the angle $\phi$ is challenging with the current data. Transiting exoplanet detections report the time of each inferior conjunction, that is, when the mean longitude $\lambda=\pi/2$. However, libration angles depend on mean longitudes at different points in the orbit. Even if the system is assumed to be coplanar, a global system fit requires 5 parameters per planet, for a total of 20 free parameters. Kepler-221 shows only weak TTVs, effectively precluding any chance to constrain planet masses and eccentricities. Therefore, actually determining whether any critical angle is librating for Kepler-221 is probably impossible with current measurements. 

\begin{figure}[h]
    \centering
    \includegraphics[width=0.5\textwidth]{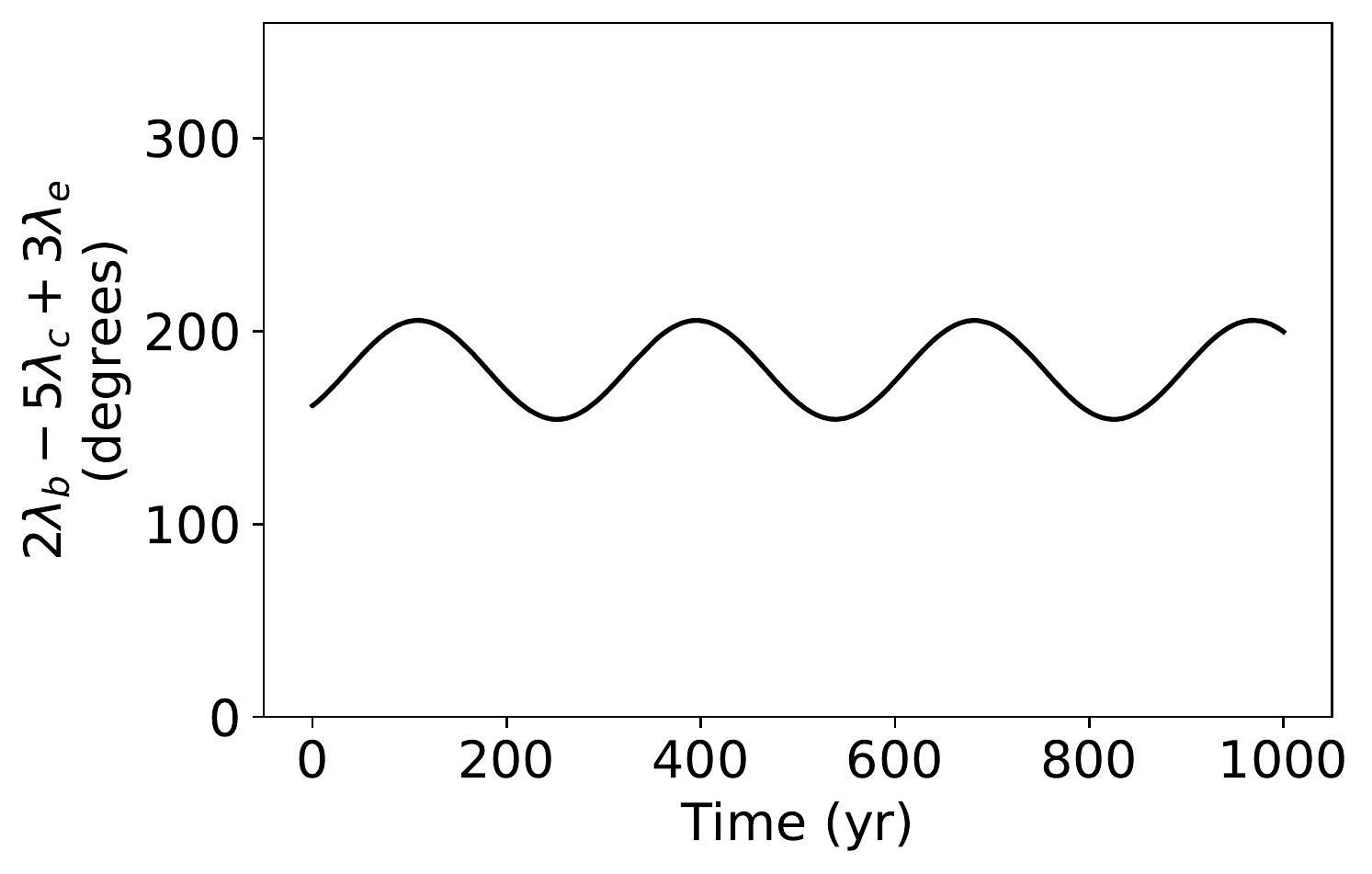}
    \caption{Libration of the three-body resonant angle in a particular configuration of Kepler-221 planets consistent with available data.}
    \label{fig:lib}
\end{figure}

However, because there are many transits, the orbital frequencies \textit{are} well-constrained and therefore the derivative $\dot{\phi} = B$ is also well-constrained. So, we appeal to the closeness of $B$ to zero to argue that the angle $\phi$ is likely librating. \added{To demonstrate, we pick a particular set of initial conditions for Kepler-221 that is consistent with TTVs (Table \ref{config}). Because the true planet masses are unknown, we chose them arbitrarily, ensuring they agree with the probabilistic mass-radius relation of \citet{Chen2017a}. For definitiveness, we assume coplanar and initially circular orbits throughout. For all integration reported in this work, we use the \texttt{whfast} integrator in the \texttt{rebound} software package and set the timestep to between $1/15$ and $1/12$ of the period of the innermost orbit \citep{Rein2015}. This set of initial conditions produces libration of the critical angle $\phi$ around $180^\circ$ (Figure \ref{fig:lib}). Furthermore, using the three-body resonance properties analytically calculated in \cite{Quillen2011}, we find the resonance width in frequency ($B$) space to be $\sim0.0004$ degrees/day, comparable to the measured value of $B$.} \deleted{As a demonstration that libration of $\phi$ is not incompatible with the data, in Figure \ref{fig:lib} we show the libration of $\phi$ using a particular set of initial conditions for Kepler-221, given in Table \ref{config}, that is consistent with TTVs.} \deleted{Because the true planet masses are unknown, we chose them arbitrarily, ensuring they agree with the probabilistic mass-radius relation of \citet{Chen2017a}. For definitiveness, we assume coplanar and initially circular orbits throughout. For all integration reported in this work, we use the \texttt{whfast} integrator in the \texttt{rebound} software package and set the timestep to between $1/15$ and $1/12$ of the period of the innermost orbit \citep{Rein2015}.}

\begin{deluxetable*}{cccc}
\tablecaption{Initial conditions used for simulations\label{config}}
\tablewidth{0pt}
\tablehead{
\colhead{Object} & \colhead{Mass ($M_\oplus$)} & \colhead{Period (d)} & \colhead{Mean longitude (deg)}
}
\startdata
    Star & 0.95 $M_\odot$ & & \\
    b & 3.0 & 2.79584 & 176.7625 \\
    c & 9.6 & 5.69023 & 67.8067 \\
    d & 9.0 & 10.04346 & 7.3874 \\
    e & 9.6 & 18.37103 & 168.9630 \\
\enddata
\end{deluxetable*}

In principle, one could argue that given the large number of Kepler multiplanet systems and the number of possible coefficients on the critical angle, the slow evolution of the critical angle for Kepler-221 is coincidental. To test this, for each transiting exoplanet system with at least 3 planets, obtained from the Exoplanet Archive,\footnote{\url{exoplanetarchive.ipac.caltech.edu}} we computed the distance to resonance parameter
\begin{equation}
    B = pn_1 - (p+q)n_2 + qn_3
\end{equation}
for all \added{positive coprime} integer values of \replaced{$p,q < 5$}{$p,q$ with $p+q<10$}, where $n_1, n_2, n_3$ are the mean motions for any three distinct planets in the system \added{in ascending period order,} resulting in approximately \replaced{20,000}{18,000} individual frequencies. We then scale this frequency by the average of the mean motions $\langle n \rangle$ to get a dimensionless frequency $|B|/\langle n \rangle$. The values of $|B|/\langle n \rangle$ closest to zero are shown in Table \ref{B} \added{along with basic information about the system and whether the planet triplet is adjacent}. TRAPPIST-1, \added{TOI-178,} Kepler-80, Kepler-60, K2-138, \replaced{Kepler-11}{K2-72}, and Kepler-223 are known resonant chain systems \citep{Gillon2017, Leleu2021, MacDonald2016, Gozdziewski2016, Christiansen2018, Migaszewski2012, Mills2016}. \deleted{The rest are several orders of magnitude larger and therefore likely far enough from a three-body resonance to have a librating resonant angle.} The fact that $|B|/\langle n \rangle$ for Kepler-221 is characteristic of resonant chain values is indicative that its closeness to zero is not coincidental. \added{We note also that the next smallest value of $B/\langle n \rangle$ for Kepler-221, $3n_c - 8n_d + 5n_e \approx 0.96$ degrees/day, is more than 3 orders of magnitude larger than the one given in Eq. \ref{eq:B}. We conclude that no further commensurabilities exist in the system, and that planet d is not involved in the resonant dynamics.}

\begin{figure}[h]
    \centering
    \includegraphics[width=0.5\textwidth]{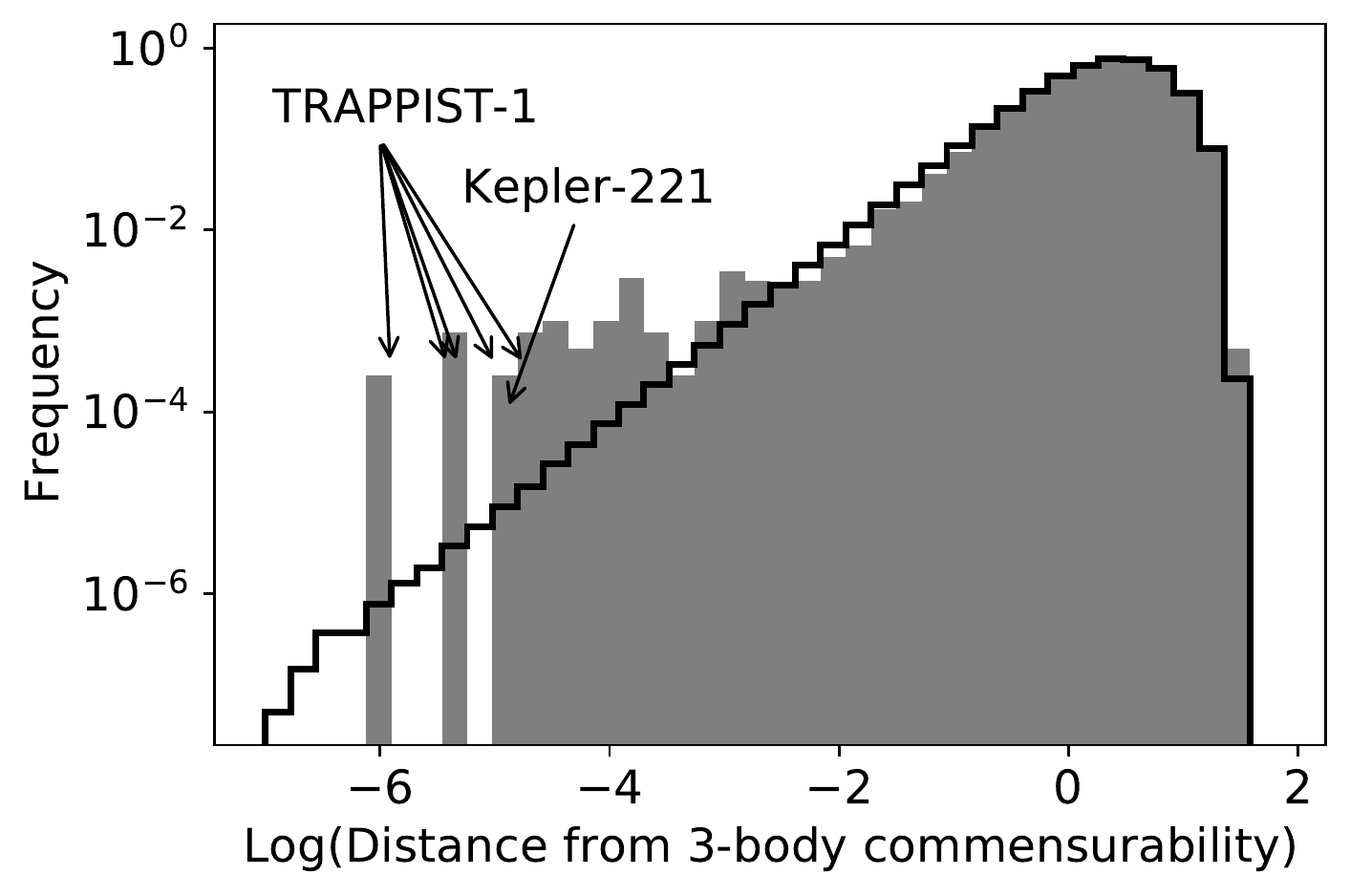}
    \caption{The distribution of the $B$ parameter normalized by average mean motion, indicating closeness to resonance. The gray histogram is the distribution of all computed $B/\langle n \rangle$ dimensionless frequencies from the transiting multiplanet sample. The black line is the simulated distribution formed by reshuffling all planet periods.}
    \label{fig:bootstrap}
\end{figure}

As a separate check, we compared the distribution of all values of $|B|/\langle n \rangle$ with one obtained by bootstrapping randomly selected periods. Specifically, for each transiting system with $N$ planets, we drew $N$ planet periods randomly from the distribution of all planet periods, without replacement, and computed the new distance to resonance parameter as above. We repeated this bootstrapping process \replaced{1000}{10000} times. Figure \ref{fig:bootstrap} shows the distribution of the dimensionless parameter $|B|/\langle n \rangle$ for the bootstrapping test compared to the observed values. There appear to be two distributions: non-resonant planet triplets are uniformly distributed near zero, but a small number of planet triplets have exceptionally small values of $|B|/\langle n \rangle$. Kepler-221 appears to be part of the latter distribution. Its lowest value of $|B|/\langle n \rangle$ is larger than only $3\times 10^{-6}$ of the bootstrap distribution. Therefore, only 0.06 of the \replaced{20,000}{$\sim18,000$} computed frequencies would be expected to be less than it. We thus conclude that Kepler-221 is unlikely to coincidentally lie at this commensurability, and note that there are no other systems with such a property that have not been previously identified as unique three-body resonant chains in the Kepler sample.

\begin{deluxetable}{ccccccc}
\tablecaption{The ten systems with frequencies $B$ closest to zero, for all transiting planet systems; we show only the frequency nearest to zero for each system.\label{B}}
\tablehead{
\colhead{System} & \colhead{$|B|$ (deg day$^{-1}$)} & \colhead{$|B|/\langle n \rangle$ (deg)}  & \colhead{B expression} & \# planets & Adjacent? & \colhead{Resonant Chain?}}
\startdata
    TRAPPIST-1 & 0.000072 & 0.000070 &   $n_d - 2n_e + n_g$ &   7 & No &  Yes \citep{Gillon2017}\\
    Kepler-221 & 0.000727 & 0.000590 & $2n_b - 5n_c + 3n_e$ &   4 & No & ? \\
    TOI-178 & 0.000649 & 0.001447 &  $n_e - 3n_f + 2n_g$ &   6 & Yes & Yes \citep{Leleu2021} \\
    Kepler-80 & 0.006714 & 0.007537 &   $n_e - 2n_b + n_g$ &   6 & No & Yes \citep{MacDonald2016} \\
    Kepler-60 & 0.005504 & 0.007814 &   $n_b - 2n_c + n_d$ &   3 &  Yes & Yes \citep{Gozdziewski2016} \\
    K2-138 & 0.008791 & 0.010918 & $2n_d - 5n_e + 3n_f$ &   5 &  Yes & Yes \citep{Christiansen2018} \\
    K2-72 & 0.030943 & 0.039504 & $5n_b - 9n_d + 4n_c$ &   4 &  Yes  & Maybe \citep{Crossfield2016} \\
    Kepler-223 & 0.035062 & 0.054959 &   $n_b - 2n_c + n_d$ &   4 &      Yes & Yes \citep{Mills2016} \\
    Kepler-327 & 0.083137 & 0.060542 & $3n_b - 8n_c + 5n_d$ &   3 &      Yes & ?\tablenotemark{a} \\
    Kepler-184 & 0.028801 & 0.077570 &  $n_b - 4n_c + 3n_d$ &   3 &      Yes & ?\tablenotemark{a}\\
\enddata
\tablenotetext{a}{We were unable to find detailed analyses of either of these systems.}
\end{deluxetable}

\section{In-situ resonance capture} \label{sec:insitucap}
If the planets of Kepler-221 are indeed entrained in a three-body resonance, an immediate question arises: how was this commensurability established? Resonances represent only a small fraction of the parameter space and are not expected to occur coincidentally. Instead, they are effective potential wells that act as attractors under convergent migration or other dissipative mechanisms \citep{Papaloizou2010}. As migrating planets pass through orbital commensurability, the resonance can ``capture,'' causing the planet pair to remain in resonance as migration continues \citep{Henrard1982, Borderies1984}. Provided sufficiently slow migration, resonance capture is highly effective, so much so that the dearth of resonances in the Kepler sample constitutes a problem \citep{Adams2008, Izidoro2017}. The case of two-body resonance capture is well-understood \citep{Batygin2015}. For pure three-body resonances, numerical simulations of migration and resonance capture of three-planet systems have found pure three-body resonances to be extremely rare or nonexistent \citep{Gallardo2016, Charalambous2018}.

\begin{figure}
    \centering
    \includegraphics[width=0.8\textwidth]{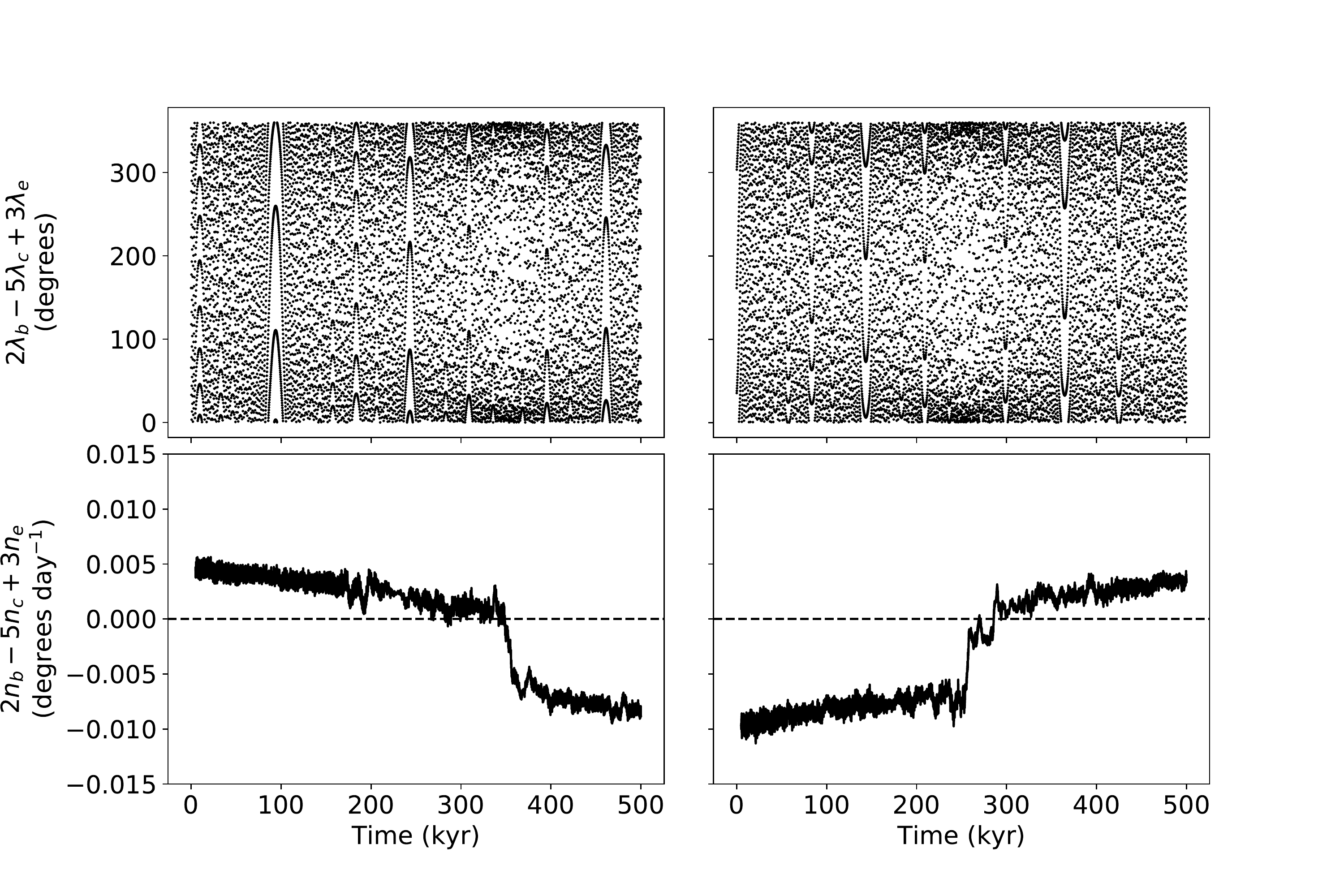}
    \caption{Encounters of the Kepler-221 planet with a pure three-body resonance by convergent (left) and divergent (right) migration of the outer planet e with 10 Gyr $e$-folding time. The top panels shows the circulation of the critical angle and the bottom panels show the frequency $B$ defined in the text. In either case, the adiabatic limit is broken, and resonant capture fails.}
    \label{fig:capture}
\end{figure}

Nevertheless, Kepler-221 might represent a pathological case, and we have attempted to simulate capture numerically. Starting with the parameters used for Figure \ref{fig:lib}, we displaced planet e slightly inside or outside of the resonance, and then applied semi-major axis migration towards the resonance. Even at the extremely slow migration rate of 10 Gyr $e$-folding time, far slower than protoplanetary disk-driven migration \citep{Paardekooper2010}, convergent and divergent migration are not sufficient to capture into a pure three-body resonance. The critical angle continues to circulate as the system passes through the resonance and exhibits the characteristic jump over the width of the resonance (Figure \ref{fig:capture}).

We also considered the probability of resonance capture from the perspective of Hamiltonian perturbation theory (Appendix \ref{hammy}). Importantly, this analysis demonstrates that the resonance widths are sufficiently small that the migration rates required for adiabatic capture are far longer than the system's age, and even in the case of adiabaticity, the capture probability is negligible. We thus conclude that direct capture of planets b, c, and e into a three-body resonance in Kepler-221 is practically impossible.

\section{Indirect Capture} \label{sec:indcap}
The results of the previous section lead us naturally to consider indirect forms of capture. In particular, we consider the possibility that libration of the 3-body angle stems from simultaneous libration of 2-body angles facilitated by rapid circulation of $\varpi_c$. Although the planets in Kepler-221 lie far from exact two-body commensurabilities, previous work \citep{Batygin2013, Pichierri2019} has shown that libration of resonant angles can be maintained well outside the nominal resonant width, as long as eccentricities are very small. In particular, c and b have period ratio 2.035, wide of a 2:1 resonance, and e and c have period ratio 3.228. There are two choices of critical angle for the 2:1 eccentricity resonance and three for the 3:1 eccentricity resonance, depending on the coefficients of the arguments of periastron. However, if we choose the critical angle
\begin{equation}
    \phi_1 = \lambda_b - 2\lambda_c + \varpi_c
\end{equation}
for the 2:1 resonance and
\begin{equation}
    \phi_2 = \lambda_c - 3\lambda_e + 2\varpi_c.
\end{equation}
for the 3:1 resonance, the linear combination $2\phi_1-\phi_2$ leads to a cancellation of the arguments of periastron and recovers the three-body critical angle $\phi$. In other words, if both two-body resonant angles librate with small amplitudes, the three-body angle will librate as well. In two-body resonances, dissipative evolution results in so-called ``resonant repulsion'' in which the period ratio increases as energy is liberated \citep{Lithwick2012, Batygin2013}. If this mechanism also operates on a resonant chain with period ratios of 1:2:6, a strong damping mechanism could push the planets far wide of their resonances \citep{Pichierri2019}. If the three-body commensurability were also preserved, it could result in a system similar to Kepler-221.

The process of building a resonant chain, characterized by libration of specific harmonics, is sensitive to initial eccentricities and inclination, migration rates, order of assembly, efficiency of eccentricity damping, etc. Ideally, here would model it with an eye towards presenting a self-consistent migration to capture to divergence scenario for Kepler-221's architecture. However, we lack even basic information about the system, such as planet masses. With our estimates, quoted in Table \ref{config}, we were unable to model this complete evolutionary history of the system because our chosen parameters did not form a 1:2:6 resonant chain under convergent migration. Thus, rather than repeatedly turning knobs, we ask: can this process work in principle? For this, we turn to a simpler toy system using a pair of 2:1 resonances, inspired by the widely known example of a resonant chain in the Galilean satellites of Jupiter. In this case the relevant two-body libration angles are
\begin{eqnarray}
\phi_1=\lambda_b - 2\lambda_c + \varpi_c \\
\phi_2=\lambda_c - 2\lambda_e + \varpi_c
\end{eqnarray}
and the three-body angle is
\begin{equation}
\phi = \phi_1-\phi_2 = \lambda_b - 3\lambda_c + 2\lambda_e.
\end{equation}
We choose a fiducial system, and allow it to form \textit{some} resonant chain. Then, we study its evolution as an analogy. 

\begin{figure}[h]
    \centering
    \includegraphics[width=0.8\textwidth]{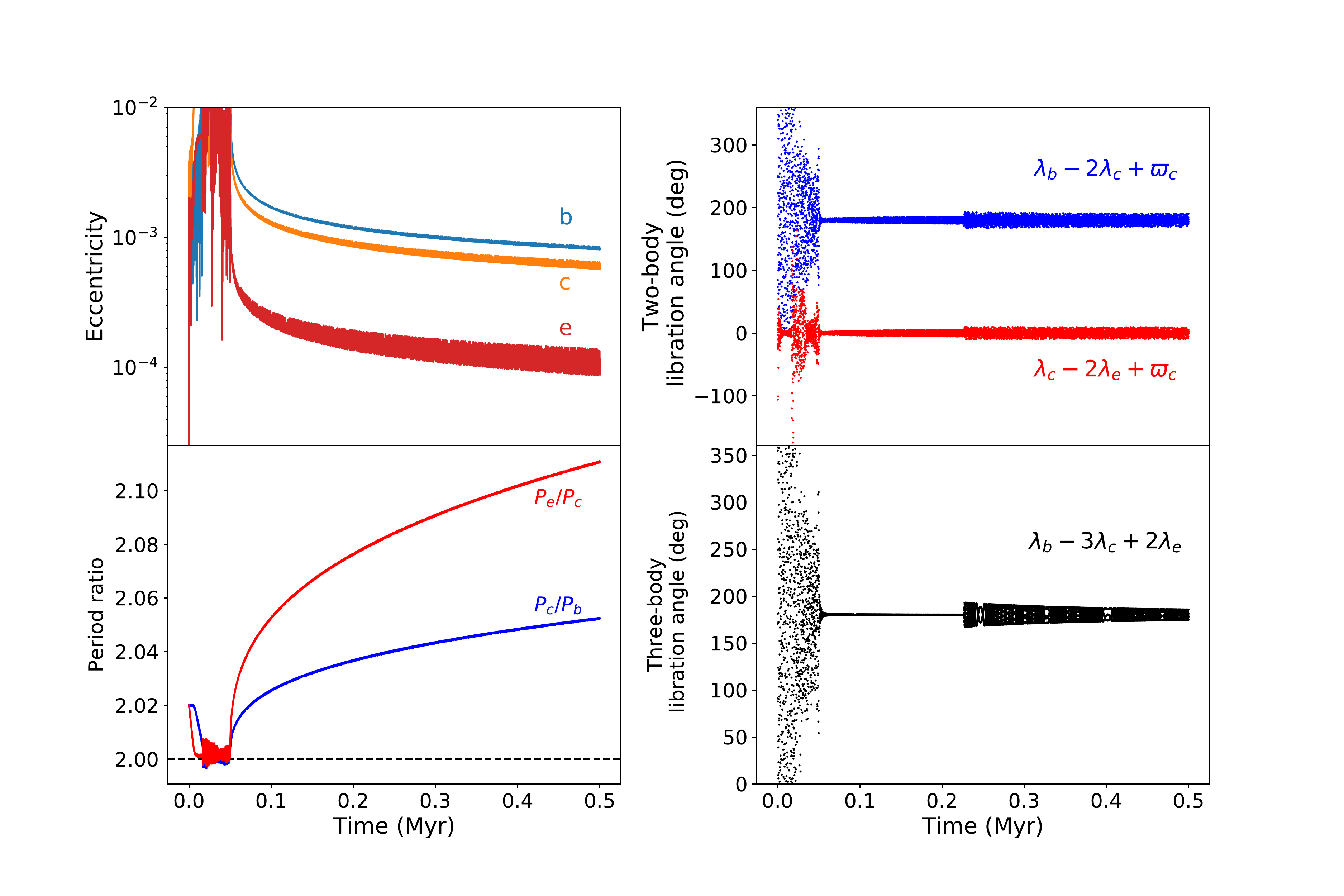}
    \caption{Toy model of formation of a Kepler-221-like system involving a pair of 2:1 resonances. Strong eccentricity damping is applied, which increases the period ratio far wide of the original resonances, nevertheless the three key resonance angles remain librating.}
    \label{fig:toy}
\end{figure}

To set up the simulation, three planets each of mass $10M_\oplus$ are added at period ratios of 2.02 around a $1 M_\odot$ star, with an initial period of 2.8d for the inner planet. We turn on convergent migration for the outer planet with migration timescale $10^6$ yr, and eccentricity damping with timescale $10^4$ yr. \added{Both effects are implemented with \texttt{reboundx} and the eccentricity damping conserves angular momentum \citep{Tamayo2020a}.} At $t = 5\times10^4$ yr, when each of the three resonant angles $\phi_1, \phi_2, \phi$ are librating, semi-major axis damping is removed and only eccentricity damping is turned on, with $\tau_e = 100$ yr for all three planets. The configuration is integrated for 0.5 Myr, by the end of which the planet pairs are at period ratios of $2.05$ and $2.12$. All three angles continue to librate. However, given the distance from resonance, the two-body angles are only formally librating, a process that ensues only at very low eccentricities driven by extremely rapid perihelion precession; the two-body resonances in this case are in \textit{forced equilibrium} \citep{Delisle2012}. The three-body commensurability is preserved as a consequence of this forced libration. Slower eccentricity damping, which is probably more realistic, could produce this configuration over longer timescales. Figure \ref{fig:toy} shows the resultant divergent migration and maintenance of libration. Even though we have not modeled Kepler-221 exactly, we believe that it would behave in the same way, given appropriate initial conditions.

Simulations of the toy model above demonstrate that dissipative evolution is confined to the line of three-body commensurability and the orbits spread wide of the original two-body resonances, analogous to the situation for two-planet systems \citep{Lithwick2012}. To confirm this mechanism with Kepler-221, we started with the current observed configuration of the system (Table \ref{config}) and applied eccentricity damping with $\tau_e = 100$ yr. Because dissipative processes are irreversible, the system will not revert to its initial state but rather continue its evolution. The results of this damping simulation, shown in Figure \ref{fig:kep221damp}, demonstrate that energy dissipation increases the period ratios of the resonant planets and preserves the three-body commensurability, even while the individual two-body resonant angles that compose it are not librating (recall that apparent libration far from the commensurability requires the system to lie exactly at the ``resonant'' focus; even a minute deviation from the global fixed point will result in apparent circulation).

\subsection{Energy evolution of Kepler-221}
\begin{figure}
    \centering
    \includegraphics[width=0.5\textwidth]{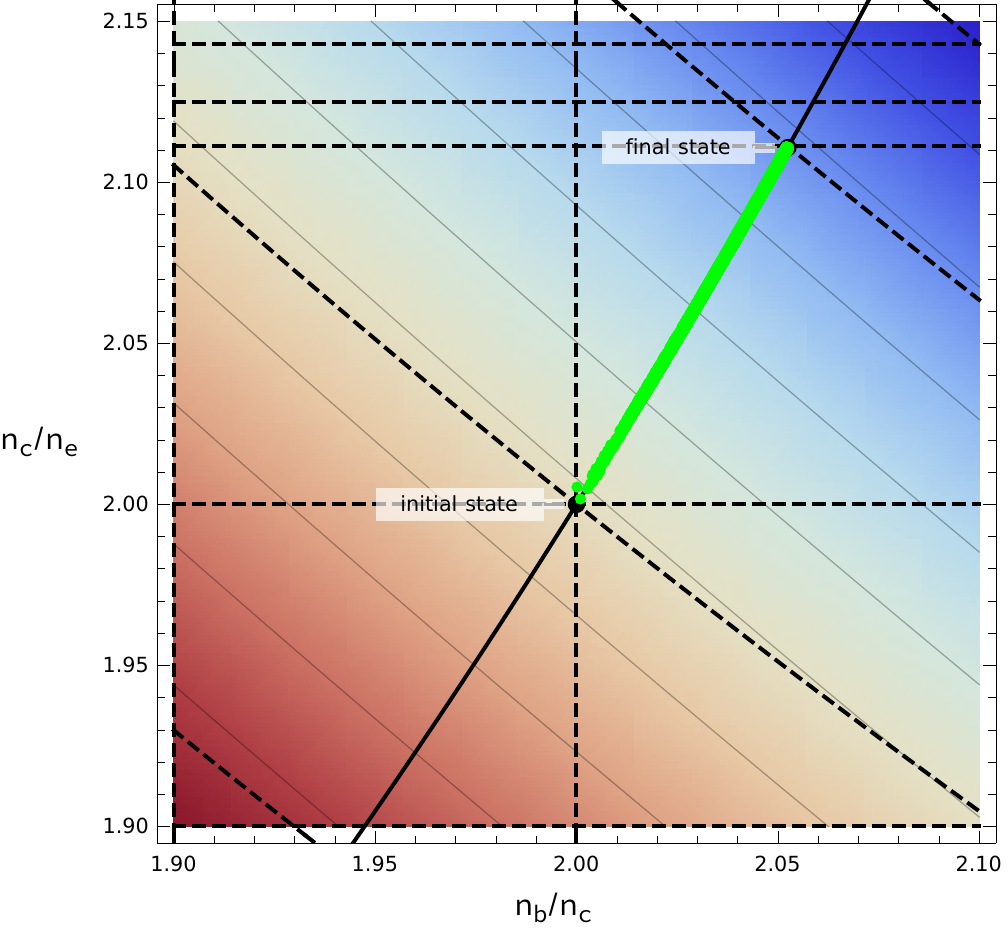}
    \caption{Period commensurabilities relevant to our fiducial model where a Kepler-221 analog is initialized in a 1:2:4 resonant chain. In these coordinates, resonances up to the tenth order between planets b and c are represented by vertical dashed lines, c and e by horizontal dashed lines, and b and e by downward-sloping dashed lines. The three-body relation $n_b-3n_c+2n_e=0$ is shown by the upward-sloping black line. Green points show the simulated evolution from Figure \ref{fig:toy}. Thin gray contours and background shading illustrate the total orbital energy, assuming constant angular momentum; energy decreases to the upper right corner. }
    \label{fig:toy_energy}
\end{figure}

The detailed evolutionary history of three-planet systems depends on many factors. However, we can understand it in relatively simple terms by noting that tidal dissipation removes energy but conserves angular momentum. Neglecting interaction terms of $\mathcal{O}(m/M_\star)$ \citep{Quillen2011}, the total orbital energy is
\begin{equation}
E = -GM_\star\left(\frac{m_b}{2a_b} + \frac{m_c}{2a_c} + \frac{m_e}{2a_e}\right).
\end{equation}
The total angular momentum, assuming coplanar and circular orbits, is
\begin{equation}
L = \sqrt{GM_\star}\left(m_b\sqrt{a_b} + m_c\sqrt{a_c} + m_e\sqrt{a_e}\right).
\end{equation}
Angular momentum conservation provides a transformation from three-dimensional semi-major axis space ($a_b, a_c, a_e$) to two-parameter period ratio space $(n_b/n_c,n_c/n_e)$. Because each point in that parameter space corresponds to three semi-major axes, we can define the scalar function $E(n_b/n_c,n_c/n_e)$. Additionally, in period ratio space, commensurabilities (two- and three-body) become one-dimensional lines. These commensurabilities, as well as the evolution our fiducial model through period ratio space, is shown in Figure \ref{fig:toy_energy}. The primary advantage of the above analysis is that is does not depend on the detailed nature of the energy dissipation. Nevertheless, we can conjecture plausible mechanisms and rule out others.

Tides on the planet generated by eccentric orbits have been shown to be insufficient at reproducing the Kepler population, within which there is an overpopulation of planet pairs $\sim 5\%$ wide of resonance \citep{Lee2013, Silburt2015}. Kepler-221 has been noted as a possible exception to this pattern, but this is because previous studies have classified planets with $R<2R_\oplus$ as ``Earth-like'' and assigned optimistic values for the dissipation parameter $k_2/Q$ of $1/100$ \citep{Lee2013} and $1/40$ \citep{Silburt2015}, in addition to assuming a solar-like age of $>5$ Gyr. Fortunately, we need not carry out the fully-fledged simulations of dissipation-driven orbital divergence, since the behavior of this process is understood specifically. Tidal dissipation has a characteristic power law growth in which the period ratio is the exact commensurability plus a term that grows as $(t/\tau_e)^{1/3}$. Using this prescription and confirming it with direct simulations, we evolved Kepler-221 including eccentricity damping with $\tau_e=100$ yr for planet b and found that approximately 7000 cycles of $\tau_e$ must have elapsed to reach the present configuration. The damping timescale associated with eccentricity tides is
\begin{equation}
\tau_e = \frac{2}{21n_b} \frac{Q}{k_2} \frac{m_b}{M} \left(\frac{a_b}{R_p}\right)^5
\end{equation}
where $k_2$ is the Love number, $Q$ is the tidal quality factor, $R_p$ is the planet radius. Assuming an optimistic system age of 1 Gyr and the parameters in Table \ref{config}, we constrain $k_2/Q \gtrsim 1/3$, i.e. an order of magnitude more dissipative than Earth and any body in the Solar System \citep{Murray1999}. Based upon this estimate, we strongly disfavor eccentricity tides as a likely mechanism for driving the long-term orbital divergence of Kepler-221's planets.

\begin{figure}
    \gridline{\fig{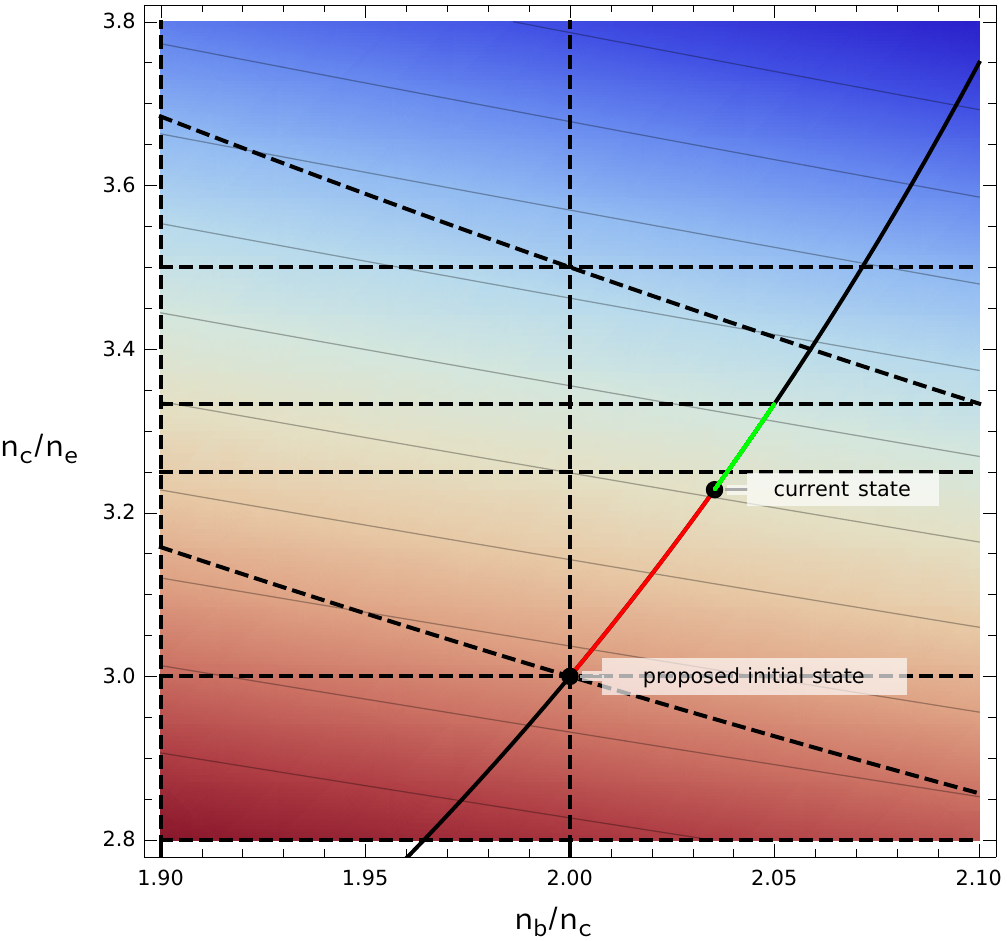}{0.4\textwidth}{(a)}
          \fig{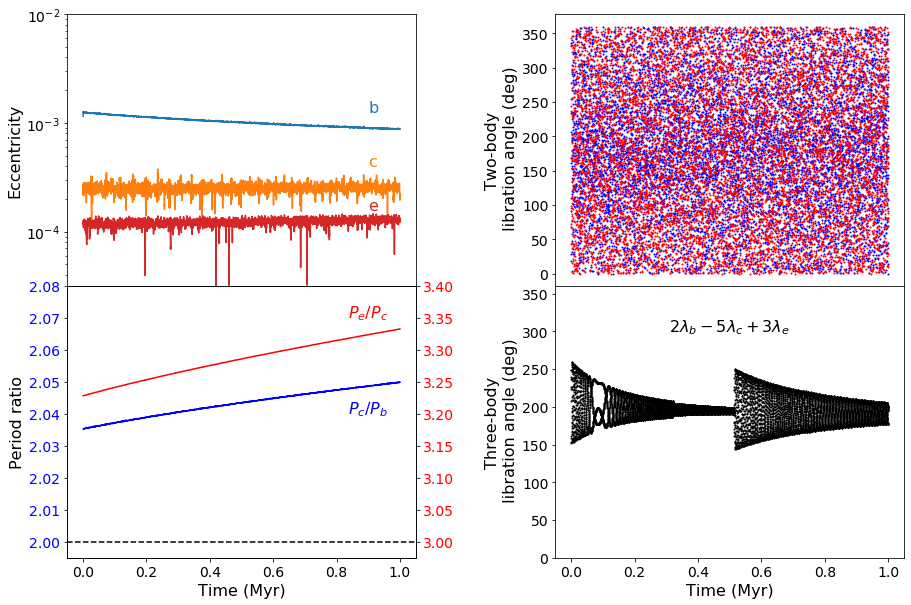}{0.55\textwidth}{(b)}}
    \caption{Evolution of the Kepler-221 system under eccentricity damping with $e$-folding time 100 yr acting on each planet, simulating energy dissipation at constant angular momentum. (a) Period commensurabilities in period-ratio space, analogous to Figure \ref{fig:toy_energy}. Here, the red line marks the evolution between the proposed initial state and the current state of Kepler-221 along the three-body resonance, and green points show the simulated continued evolution. (b) Period ratios, eccentricity, and resonant angles during the simulated dissipation. The uniform scatter in the two-body resonance angles indicates circulation.}
    \label{fig:kep221damp}
\end{figure}

Thankfully, there are more efficient variants of tidal dissipation. One possibility is obliquity tides, which can be maintained over long timescales by a secular spin-orbit resonance and require mutual inclination \citep{Millholland2019}. Kepler-221 is an inclined system, with a minimum mutual inclination of $\sim 7^\circ$, making it an excellent candidate for damping by obliquity tides. We can estimate the dissipation provided by obliquity tides as a feasibility check. Assuming circular orbits, and equilibrium rotation, each planet removes energy at the rate
\begin{equation}
\frac{dE}{dt} = \frac{2\sin^2\epsilon}{1+\cos^2 \epsilon} \frac{3n}{2} \frac{k_2}{Q} \left(\frac{GM^2}{R_p}\right) \left(\frac{R_p}{a}\right)^6
\end{equation}
where $\epsilon$ is the obliquity of the planet \citep{Millholland2019}. For typical obliquities driven by the secular spin-orbit resonance the first factor is approximately unity. Taking our optimistic system age of 1 Gyr, the values of $k_2/Q$ needed to damp the energy from the initial to final state are roughly $1/100,000$, $1/50,000$, and $1/100$ assuming damping by only b, c, or e, respectively. The first two values are plausible for super-Earths or sub-Neptunes \citep{Morley2017, Puranam2018}, while the last is perhaps too high, although considerable uncertainties exist \citep{Efroimsky2007}.
Hence, the presence of either planet b or c in an excited obliquity state is sufficient to damp the system to its current state.

Note that in all of these situations, if the migrating planets encounter even a weak two-body resonance, the libration of the three-body angle may break. This is likely because as planets pass through resonance (but do not capture), they experience a jump in semi-major axis. Even a small jump is enough to escape the extremely narrow three-body libration width. For example, in one integration, $\phi$ began circulating when the outer planets passed through a period ratio of 10:3. The near-instantaneous increase in libration amplitude in Figure \ref{fig:kep221damp} is also likely due to an encounter with a very high order resonance. Therefore, a necessary condition for the initial and final state of a system like Kepler-221 is that there is no remotely strong intervening two-body resonance. An initial state corresponding to 2:1 and 3:1 resonances fulfills this criterion, as there are no two-body resonances of order $\leq 10$ and no zeroth-order three-body resonances with $p,q \leq 10$ in between the 1:2:6 initial state and the current period ratios of Kepler-221 (Figure \ref{fig:kep221damp}).

\subsection{Radius Inflation}
If indeed Kepler-221b is experiencing a large amount of tidal dissipation, the heat flux may affect the atmosphere. Here, we build a simplified atmospheric model to show that even a low-mass envelope can expand to a large size and cause a super-Earth type planet to lie within the radius gap. 
The model we aim to construct is merely illustrative. Thus, for definitions, we assume that the atmosphere is purely hydrogen, is in hydrostatic equilibrium, and obeys a polytropic equation of state, $P=k \rho^{1+\gamma}$ with $\gamma=7/5$, implying a nearly fully convective envelope. We further assume the ideal gas law with an equilibrium temperature of 1120 K, computed with an albedo of 0.3. The energy transfer equation at the radiative-convective boundary $R_{RCB}$ is
\begin{equation}
\frac{L}{4\pi R^2_{RCB}} = \frac{16}{3} \left(\frac{\sigma T^3}{\kappa \rho_{RCB}}\right) \frac{dT}{dr}
\end{equation}
where $L$ is the internal luminosity, $\sigma$ is the Stefan-Boltzmann constant, $\kappa$ is the opacity (for which we use 0.1 cm$^2/$g), and $dT/dr\approx g/c_P$ is the adiabatic lapse rate, where $g$ is the surface gravity and $c_P\approx 7R/2$ is the specific heat capacity at constant pressure. 

\begin{figure}[h]
    \centering
    \includegraphics[width=0.6\textwidth]{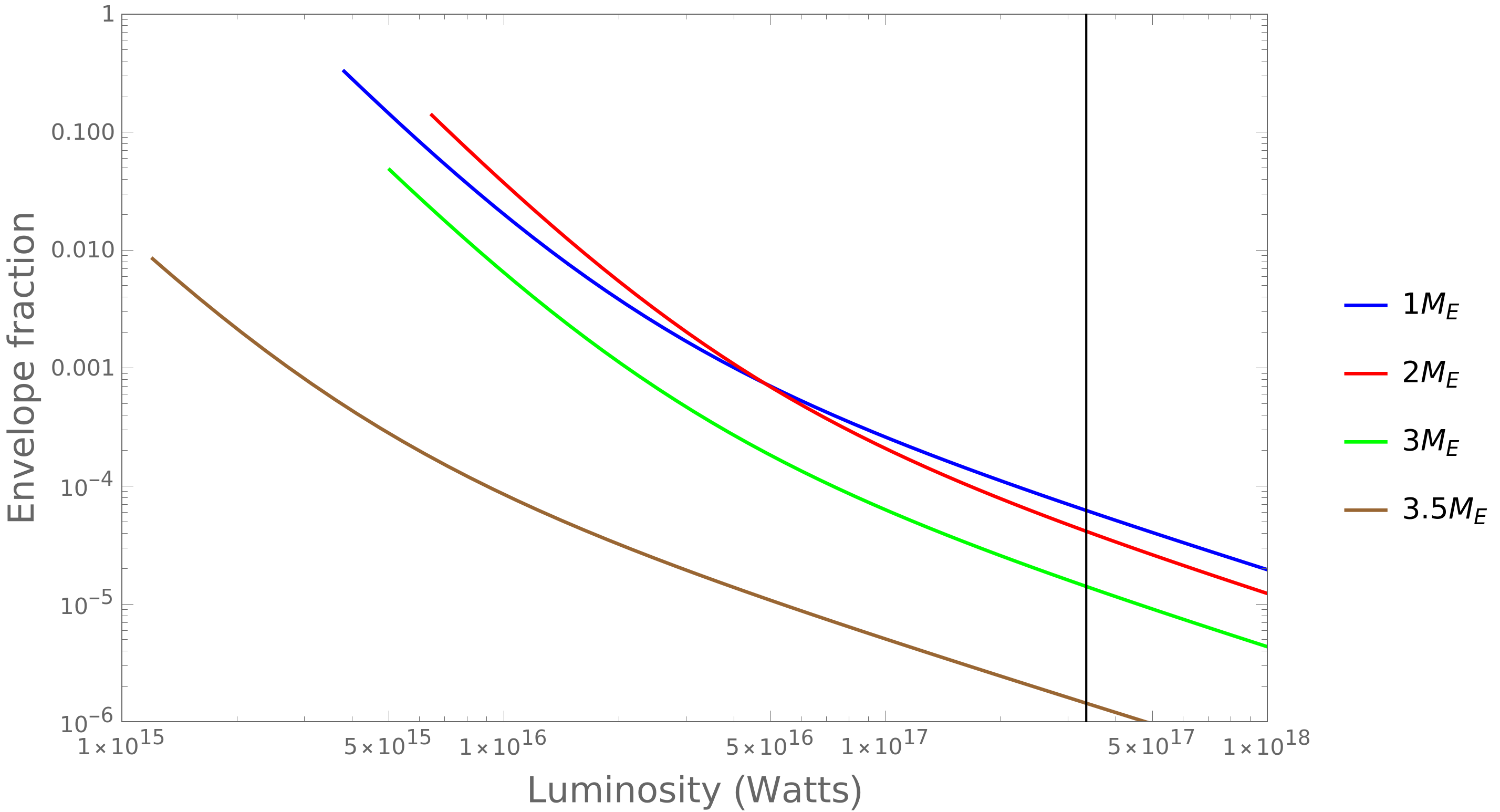}
    \caption{The mass fraction of the hydrogen envelope needed to produce a planet of radius $1.71 R_E$ as a function of tidal luminosity, for four different core masses. Each line begins at the luminosity from the heat of formation; the black vertical line is the estimated tidal luminosity of Kepler-221b.}
    \label{fig:radius}
\end{figure}

We then solve the hydrostatic equation for a variety of core masses and luminosities to obtain $\rho(r)$. Assuming the atmosphere sits atop a solid core of mass $M_c$, density 4 g/cm$^3$, and radius $R_c$, the atmospheric mass is
\begin{equation}
M_\text{env} = \int_{R_c}^{1.7 R_E} 4\pi \rho(r) r^2 dr.
\end{equation}
The atmospheric mass relative to the mass of the core of the planet is shown in Figure \ref{fig:radius}. We assume the luminosity is greater than that due to the heat of formation,
\begin{equation}
L_\text{min} \approx \frac{GM_\text{env}M_c}{(1.7 R_E)^2 \mathcal{T}}
\end{equation}
where $\mathcal{T}\approx 600$ Myr is the age of the system. If the luminosity is indeed near $3\times 10^{13}$ W as we predict, envelope fractions as small as $10^{-4}$ or $10^{-5}$ would be sufficient to yield a total radius of $1.71 R_E$. 

One hypothesis to explain the radius gap is photoevaporation \citep{Fulton2017}. In this scenario, FUV flux in the early life of the star blows off the tenuously-held atmosphere, leaving behind bare cores. In the case of Kepler-221b, however, a truly bare core is unlikely. Outgassing (which is likely to be enhanced by internal heating) provides a continuous supply of gas which will promptly be inflated to a large radius. Estimating the degree of outgassing is outside the scope of this paper because it depends strongly on the mass and composition of the planet, but we emphasize that an atmospheric mass of only $\sim 10^{-4} M_\oplus$ is sufficient.

\section{Discussion} \label{sec:disc}
In this work, we have considered the origins of the remarkable orbital architecture of Kepler-221, and have argued that a sensible evolutionary history for this system is one where the planets originated in a resonant chain and experienced long-range orbital divergence, thanks to the action of persistent tidal damping. Moreover, our analysis points specifically to obliquity tides as the primary dissipative process at play.

Although we have presented a general overview of a mechanism to produce Kepler-221, our model is incomplete. A detailed description would require knowledge of the planets' masses, which critically affect resonant dynamics, as well as mutual inclinations, which are necessary for the spin-orbit resonances that drive obliquity tides. We leave these issues to future work, as the system architecture comes into sharper focus.

Previous studies of the role of tidal dissipation in shaping multiplanet systems have discussed the statistical imprint of different dissipation mechanisms on planetary architectures. Typically, they assume that near resonant planet pairs began in exact commensurabilty and migrated outwards \citep{Lee2013, Silburt2015, Millholland2019}. While this picture can explain planet pairs wide of resonance (provided sufficient dissipation) it is possible that some of the planetary systems formed in place and found the resonant equilibrium through eccentricity damping \citep{Pichierri2019}. However, systems with complex overlapping resonances, and in particular Kepler-221, require convergent migration and assembly at exact commensurability before divergent migration driven by dissipation, and therefore provide stronger constraints on tidal mechanisms.

If tides in one of the planets is responsible for the energy dissipation in Kepler-221, and that dissipation is still occurring, it would provide a significant internal energy source for that planet and inflate its radius. Indeed, an unusual degree of dissipation could explain why planet b has a radius lying within the radius gap of \cite{Fulton2017}, typically assumed to be carved by photoevaporation. Confirmation of this hypothesis would likely require precise mass measurements of Kepler-221b because the degree of inflation depends strongly on the envelope fraction of the planet \citep{Millholland2019a}. Nevertheless, the model outlined in this work presents a testable framework for understanding the anomalous architecture of the Kepler-221 planetary system.

\acknowledgements
We would like to thank Andrew Howard, Juliette Becker, and Daniel Fabrycky for illuminating discussions. This investigation was started as a project for a course in Exoplanet Dynamics at the University of Chicago. KB is grateful to the David and Lucile Packard Foundation and the Alfred P. Sloan Foundation for their generous support.

\appendix
\section{Hamiltonian Capture Probability}
\label{hammy}
A Hamiltonian prescription lends itself well to studying capture into mean motion resonances via migration. The Hamiltonian for a pure three-body resonance was derived by \cite{Quillen2011}. For details we direct the reader to that work, but we will simply copy the relevant formulae here. For generality, we use $i,j,k$ as labels for the three planets. As mentioned in the main text, the relevant critical angle is 
\begin{equation}
    \phi = p\lambda_i - (p+q)\lambda_j + q\lambda_k
\end{equation}
and its conjugate momentum is simply $J=\Lambda_i/p$, where $\Lambda_i=m_i\sqrt{GMa_i}$. The Hamiltonian is most simply expressed by defining $J\equiv J_0 + I$ and expanding about $J_0$. Then, it takes the form of a pendulum,
\begin{equation}
    \mathcal{H}(I,\phi) = \frac{1}{2} A I^2 + BI + \epsilon_{pq} \cos\phi
\end{equation}
where, as before,
\begin{equation}
    B = pn_i - (p+q)n_j + qn_k,
\end{equation}
and
\begin{equation}
    A = -3\left(\frac{p^2}{m_ia_i^2} + \frac{(p+q)^2}{m_ja_j^2} + \frac{q^2}{m_ka_k^2}\right)
\end{equation}
\begin{eqnarray}
    \epsilon_{pq} \approx \frac{m_im_jm_k^3}{\Lambda_k^2}\left[\frac{3n_j^2}{2}\left(\frac{1}{2n_{ij}n_{jk}} + \frac{p}{qn_{ij}^2} +
    \frac{q}{pn_{jk}^2}\right)b_{1/2}^p(\alpha_{ij})b_{1/2}^q(\alpha_{jk}) \right.\\ 
    \nonumber
    + \left(\frac{n_j}{n_{jk}} + \frac{qn_j}{pn_{ij}}\right)b_{1/2}^q(\alpha_{jk})(1+\alpha_{ij}D_\alpha)b_{1/2}^p(\alpha_{ij})\\ 
    \nonumber
    + \left.\left(\frac{n_j}{n_{ij}} + \frac{pn_j}{qn_{jk}}\right)b_{1/2}^p(\alpha_{ij})\alpha_{jk}D_\alpha b_{1/2}^q(\alpha_{jk})\right].
\end{eqnarray}
Here, $n_i,n_j,n_k$ are the mean motions, and $n_{ij}=n_i-n_j$ and $n_{jk} = n_j - n_k$. Also, $\alpha_{ij}=\frac{a_i}{a_j}$ and $\alpha_{jk}=\frac{a_j}{a_k}$, $b_{1/2}^p$ is the Laplace coefficient and $D_\alpha$ is the derivative with respect to $\alpha$. We note that the exponential approximations for $b_{1/2}^p$ provided in \cite{Quillen2011} are not sufficient for Kepler-221 because the values of $\alpha_{ij}$ are too far from 1. Exact expressions involving elliptic integrals are tractable in this low-order case.

Some key properties of the resonance can now be estimated. The libration frequency is
\begin{equation}
    \omega_{pq} \sim \sqrt{\epsilon_{pq} A}
\end{equation}
and the width of the resonance \added{in semi-major axis space} is
\begin{equation}
    \Delta a \sim \frac{4p}{m}\sqrt{\frac{2\epsilon_{pq}a}{A}} \propto ma
\end{equation}
\added{where $m$ is the planet-star mass ratio.} Note, in contrast, the width of a first-order two-body (eccentricity) resonance, such as the one introduced in Section 1, is $\Delta a \propto \sqrt{m}a$.

\begin{figure}[h]
    \centering
    \includegraphics[width=0.6\textwidth]{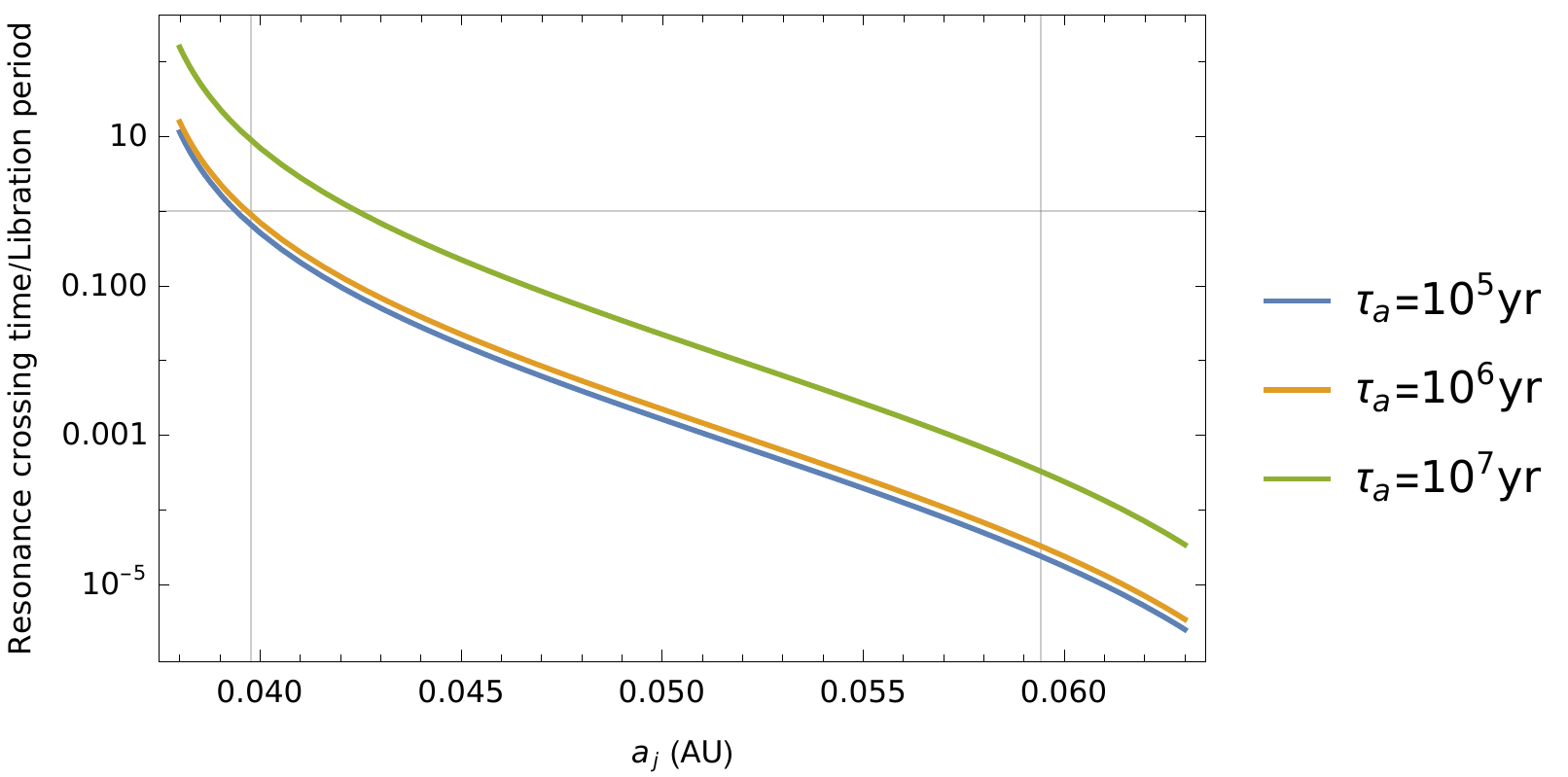}
    \caption{Adiabatic criterion (resonance crossing time/libration period) as a function of semi-major axis of the second planet in Kepler-221. The left vertical gray line marks a 5 Hill radius spacing, inside of which stability is unlikely. The right vertical gray line marks the current position of Kepler-221c.}
    \label{fig:adiab}
\end{figure}

\begin{figure}[h]
    \centering
    \includegraphics[width=0.5\textwidth]{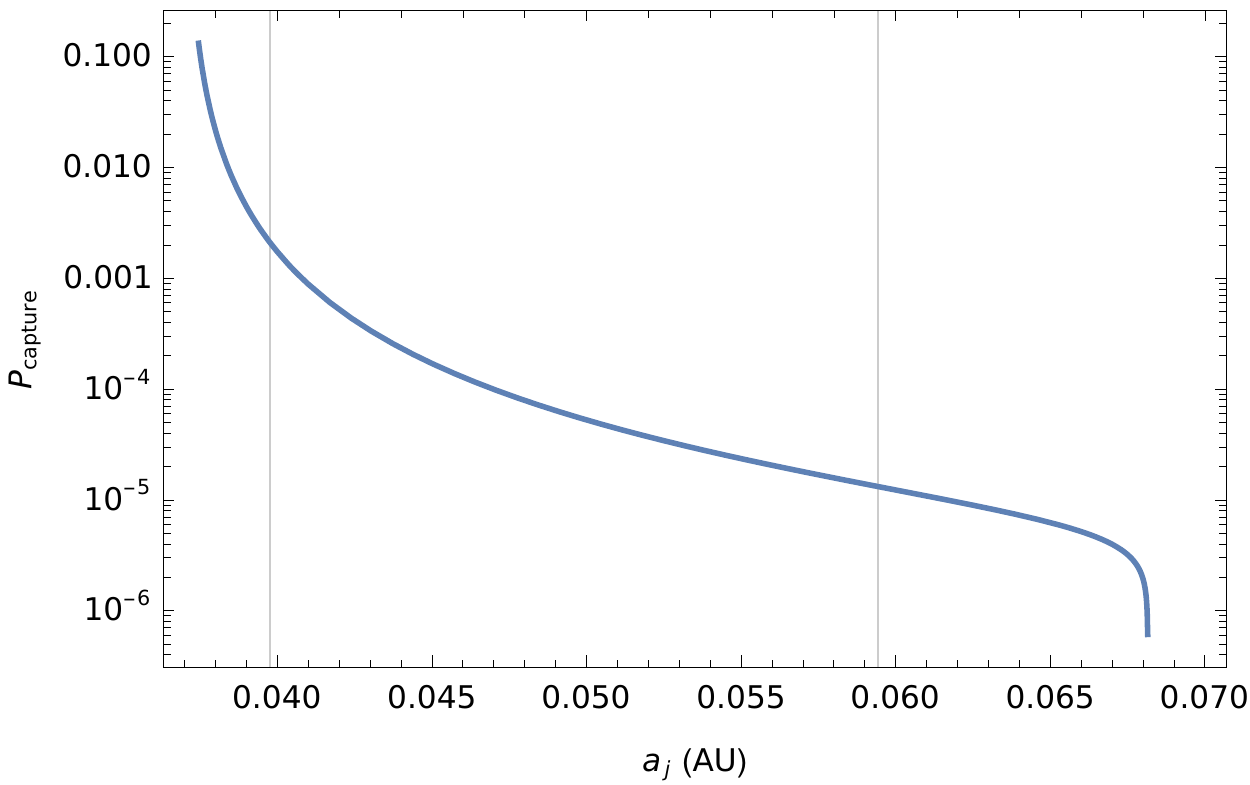}
    \caption{Probability of capture into a pure three-body resonance, assuming the adiabatic capture theory. Vertical gray lines mark the same as in Figure \ref{fig:adiab}.}
    \label{fig:pcap}
\end{figure}

Resonance capture is much more likely if the migration is adiabatic. A reasonable criterion for this is that the timescale of migration across the resonance should exceed the timescale of libration. Consider a planet migrating with semi-major axis $e$-folding time $\tau_a$. Then, the timescale for crossing the resonance is
\begin{equation}
    \Delta t = \frac{\Delta a}{\dot{a}} = \frac{\tau_a \Delta a}{a}.
\end{equation}
The adiabatic criterion is therefore
\begin{equation}
    \frac{\Delta t}{P_{\text{lib}}} = \frac{\tau_a \Delta a \omega_{pq}}{2\pi a} \gg 1.
\end{equation}
Figure \ref{fig:adiab} shows this criterion for estimated values of the Kepler-221 system. Typical migration rates within a protoplanetary disk are $\sim 10^5$ yr or less. Hence, the crossing of this three-body resonance in that scenario could not feasibly have happened adiabatically for the Kepler-221 system due to the narrowness of the resonance. There are mechanisms in which the planets could migrate much more slowly. Even then, the capture rates, as estimated from adiabatic capture theory \citep{Henrard1982}, are vanishingly low, as shown in Figure \ref{fig:pcap}. 

\newpage
\bibliography{Migration}
\bibliographystyle{aasjournal}

\end{document}